\documentclass[prl,twocolumn,aps,showpacs]{revtex4}
\usepackage{graphicx}

\begin{document}

\title{Enhanced Orbital Degeneracy in Momentum Space for LaOFeAs}

\author{Hai-Jun Zhang, Gang Xu, Xi Dai and Zhong Fang}

\affiliation{Beijing National Laboratory for Condensed Matter Physics,
and Institute of Physics, Chinese Academy of Sciences, Beijing 100190,
China;} \date{\today}

\begin{abstract}
The Fermi surfaces (FS) of LaOFeAs (in $k_z$=0 plane) consist of
two hole-type circles around $\Gamma$ point, which do not touch
each other, and two electron-type co-centered ellipses around M
point, which are degenerate along the M-X line. By
first-principles calculations, here we show that additional
degeneracy exists for the two electron-type FS, and the crucial
role of F-doping and pressure is to enhance this orbital
degeneracy. It is suggested that the inter-orbital fluctuation is
the key point to understand the unconventional superconductivity
in these materials.
\end{abstract}

\maketitle

%

The discovery of superconductivity in LaOFeAs family of compounds is
really exciting. Very soon after the discovery of up to 26 K
superconductivity in La(O$_{1-x}$F$_x$)FeAs~\cite{LOFS}, the
$T_c$=43 K in Sm(O$_{1-x} $F$_x$)FeAs~\cite{SOFS}, $T_c$=41 K in
Ce(O$_{1-x}$F$_x$)FeAs~\cite{COFS} and $T_c$=52 K in
Pr{O$_{1-x}$F$_x$)FeAs~\cite{POFS} were reported. More and more
experimental and theoretical studies suggest that the
superconductivity found here is unconventional non-BCS
type~\cite{Wang,Wen,phonon}. Several possible pairing mechanisms
have been discussed theoretically~\cite{Mazin,Kuroki}, however we
show here, based on first-principles calculations, that the key is
the inter-orbital fluctuation due to orbital degeneracy at the two
electron-type Fermi surfaces, which is much enhanced by either
F-doping or pressure.

The electronic structure~\cite{LOFP-cal,Singh,fang,kotliar} of
LaOFeAs is quasi two dimensional and very similar to typical
semi-metal. Namely there are three hole-type Fermi surfaces (FS)
around the $\Gamma $-Z line, and two electron-type FS around the M-A
line of the Brillouin zone (BZ). The existence of five FS makes the
problem complicated, however the electronic structure around Fermi
level can be qualitatively understood in a simple way using three
orbitals per Fe site ($d_{xy}$, $d_{zx}$, $d_{yz}$). Here we define
the local coordinates $x$ and $y$ to be 45$^{\circ }$ rotated from
the $a$ and $b$ directions of the crystal lattice respectively.

Let's focus on the $k_{z}$=0 plane. There are two co-centered
circles around the $\Gamma $ point, which are hole-type FS and
mostly come from the $d_{yz}$ and $d_{zx}$ states of Fe. In
addition, there are two co-centered ellipses around the M point,
which are electron-type FS and mostly come from the $d_{xy}$ and
$d_{yz/zx}$ states of Fe (see Fig.1). The anisotropy (or the shape)
of the ellipses is related to the hybridization between $d_{xy}$ and
$d_{yz}$ (or $d_{zx}$). The stronger the hybridization, the more
distortion of the ellipses. The projected orbital characters of the
electron-type FS are shown in Fig.1.  One important fact to be noted
here is that the two hole-type circles around $\Gamma $ do not touch
each other, however, the two electron-type ellipses around M are
degenerate along the X-M line (see Fig.1), and this degeneracy is
protected by the crystal symmetry of the system. However, we will
show here that additional degeneracy exists for the two branches of
electron-type FS, and the crucial roles of F-doping or pressure is
to enhance this degeneracy.

\begin{figure}[tbp]
\includegraphics[clip,scale=0.3]{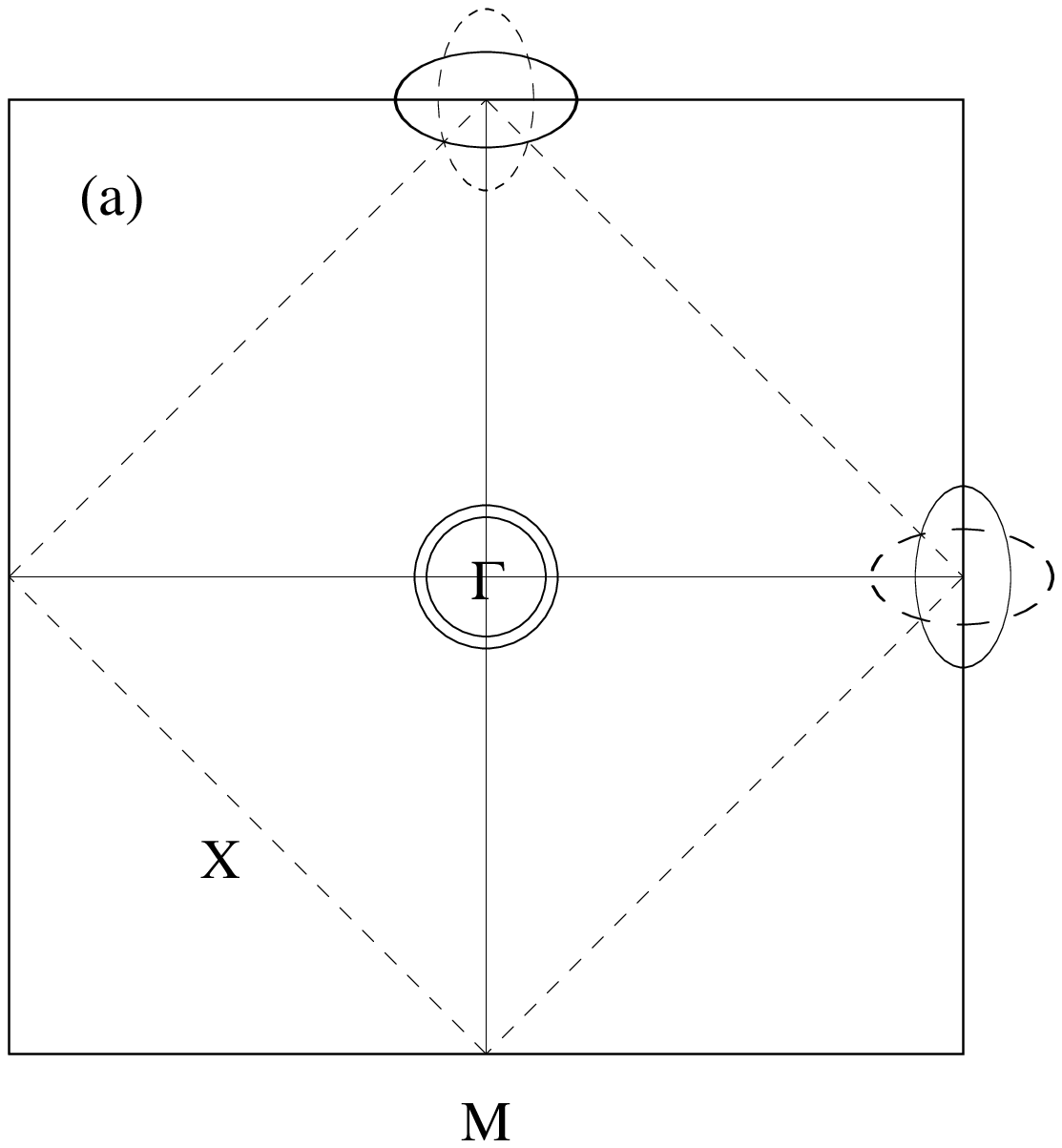} \ \ \
\includegraphics[clip,scale=0.3]{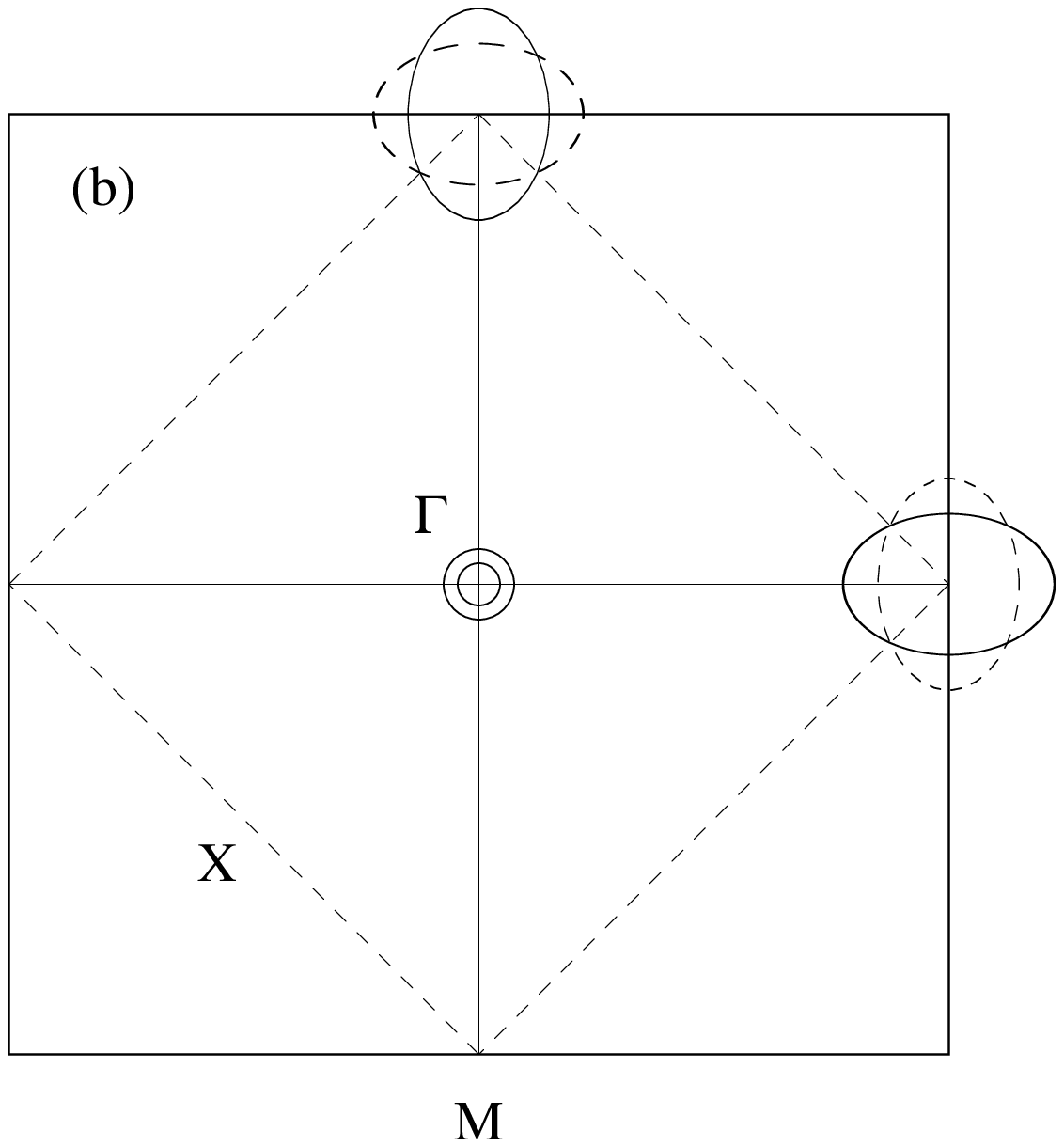}
\includegraphics[clip,scale=0.3]{fs-1c.eps}
\caption{Upper panels: Schematic plots for the Fermi surfaces of
LaOFeAs. The left panel is for pure LaOFeAs, while the right panel is
for the case with about 20\% F-doping. The large square (solid lines)
is the BZ of one Fe cell, while the small diamond (dashed lines) is
the BZ of two Fe cell. Please note the interchange of the orbital
characters of two electron-type FS after F-doping. Lower panel: The
projected orbital characters of states at FS, which is defined as
$|\langle\alpha|\Psi_k\rangle|$. Here $\alpha$ are five local Fe-3$d$
orbitals, $\Psi_k$ are Bloch states around one of the ellipses, and
angle $\theta$ is defined in the inset.}
\end{figure}

\begin{figure}[tbp]
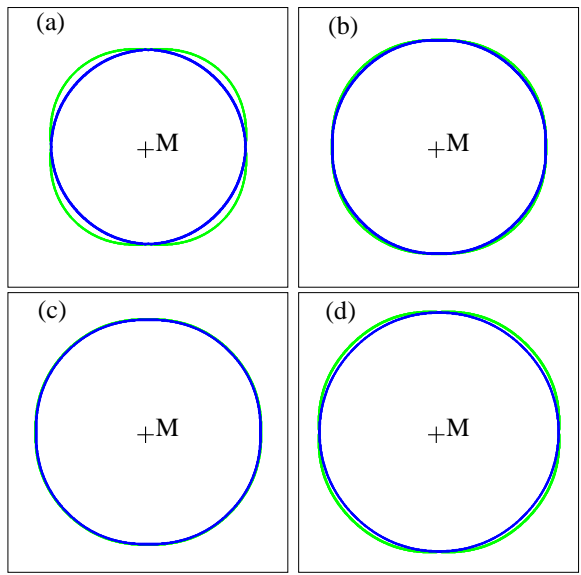

\includegraphics[clip,scale=0.23]{fig2a.eps}
\includegraphics[clip,scale=0.23]{fig2b.eps}
\includegraphics[clip,scale=0.23]{fig2c.eps}
\includegraphics[clip,scale=0.23]{fig2d.eps}
\caption{Calculated Fermi surfaces of La(O$_{1-x}$F$_x$)FeAs around the M
point for different F-doping levels. Only part of the BZ is show and the
hole-type Fermi surfaces around $\Gamma$ point is not included. From top to
bottom are, (a) no F-doping; (b) 5\% F-doping; (c) 10\% F-doping; (d) 15\%
F-doping. The two ellipses are almost degenerate for 10\% F-doping.}
\end{figure}

As discussed recently by several papers~\cite{Mazin,SDW},
significant FS nesting exists between the hole and the electron type
FS for pure LaOFeAs. After shifting the hole FS around $\Gamma$ by a
vector $q$=($\pi$, $\pi$,0), it will largely overlap with the
electron FS around M. This leads to spin-density-wave (SDW)
instability, and is the origin of the 150 K anomaly observed by
transport, susceptibility, and optical measurements~\cite{SDW}. It
is now realized that the SDW instability actually competes with
superconductivity, and super-conducting state can be realized only
after the SDW transition is suppressed~\cite{SDW,SOFS}. Electron
doping by F is an efficient way to suppress the nesting effect (the
SDW instability), because the shifting of Fermi level enhances the
mismatch between electron and hole FS (as shown in Fig.1). It is
also true that the superconductivity was observed in
LaONiAs~\cite{LONS}, where nesting effect discussed above is
irrelevant. All those information suggest that the nesting between
the electron and hole FS can not be the key issue for the
superconductivity.

Any role else played by F-doping in achieving high $T_c$
superconductivity? This is the main point we would like to address
here, based on quantitative first-principles calculations. We use
plane wave pseudo-potential method and the generalized gradient
approximation (GGA) for exchange-correlation potential, the same
approach as we reported before~\cite{fang}. As discussed above, two
hole-type FS are co-centered circles around $\Gamma$ point. They
have different size and do not touch each other. However, two
electron-type FS are ellipses around M point. They have the same
size, but are rotated by 90$^\circ$ from each other. As the result,
they overlap (or degenerate) at special k-points along the X-M line
of the BZ (see Fig.1 for schematic understanding). The size of the
electron-type FS will increase with F-doping, however, most
interestingly the shape of two ellipses is also significantly
changed by F-doping, as shown in Fig.2. With the increment of
F-doping, the anisotropy of ellipses is suppressed (i.e, they are
more and more circle-like), as the result the orbital degeneracy
right at the two electron-type FS is enhanced. By about 10\%
F-doping, the two ellipses become very close to circles and almost
exactly overlap. However, with further F-doping, two ellipses are
distorted again. 10\%-F is about the optimal doping level found
experimentally for this family of compounds, and our calculations
show that at this doping level the orbital degeneracy in the
momentum space is the highest.

\begin{figure}[tbp]
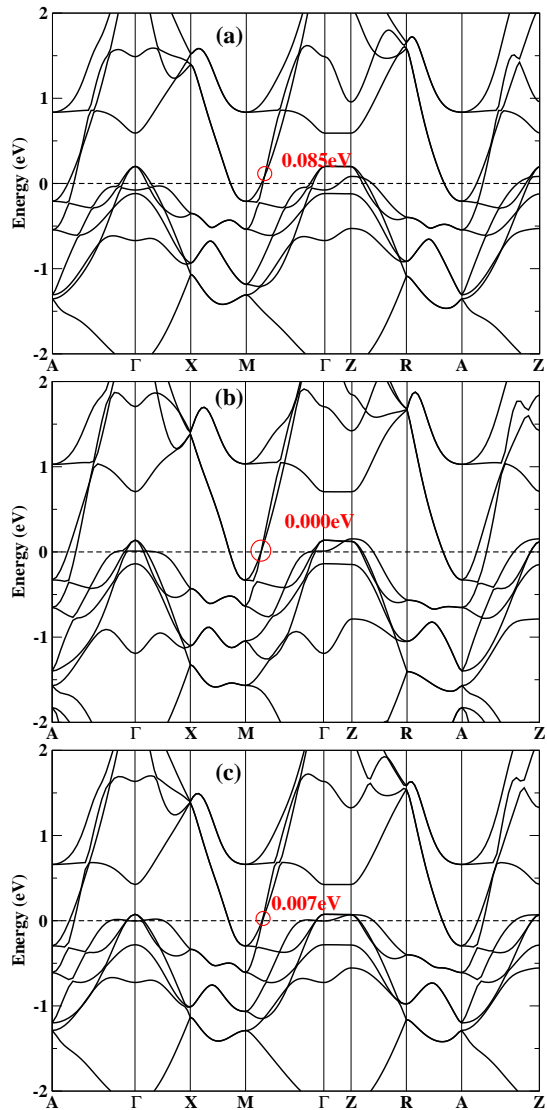

\includegraphics[clip,scale=0.3]{fig3a.eps} %
\includegraphics[clip,scale=0.3]{fig3b.eps} %
\includegraphics[clip,scale=0.3]{fig3c.eps}
\caption{Calculated band structure of LaOFeAs for (a) zero pressure; (b) 30
GPa pressure; (c) fixed volume but reducing the $c/a$ ratio by 5\%. The
crossing of two eletron-type bands is indicated by red circle.}
\end{figure}

More careful analysis about the band structure near the FS show that
additional degeneracy exists at the two electron-type FS, which can
be tuned by F-doping. This degeneracy can be understood from the
calculated band structure as shown in Fig.3(a). Looking at the
M-$\Gamma$ direct, there are two dispersive electron-type bands
crossing the Fermi level. However, the two bands disperses at
different velocity and they cross each other at about 85 meV above
the Fermi level. It is this band-crossing which leads to the
enhancement of FS degeneracy discussed above. This band-crossing is
accidental, but it is guaranteed by the different dispersion
velocities of the two bands, and it is a common feature as seen in
this family of compounds~\cite{fang}. One of the bands has mainly
character of two-dimensional $d_{xy}$ orbital, which is strongly
dispersive in the plane; another one of the bands is the linear
combination of one-dimensional orbitals $d_{yz}$ and $d_{zx}$
orbitals. The key point is that the low-lying $d_{xy}$ band (at M
point) disperses faster than the high-lying $d_{yz/zx}$ band. By
up-shift of the Fermi level (F-doping), the two electron-type FS
will become more and more degenerate, however beyond 85 meV above
the Fermi level, the two FS interchange the orbital character and
start to deviate from each other (as shown in Fig.1 and Fig.2).

In the following we discuss the effect of pressure and point out
that pressure can play a similar role as F-doping. For this
purpose, we fully optimize the crystal structure (including
internal coordinates) using the experimental space group $P4/nmm$
for each fixed volume. Then the electronic structures are
calculated using the optimized structure. Fig.4 shows the
calculated total energies, $c/a$ ratios, and internal coordinates
as function of volume. The general tendency of lattice behavior
under pressure is quite smooth and the calculated bulk modulus is
about 90 Gpa, suggesting that the compound is relatively easy to
be compressed. Other details about the effect of pressure will be
discussed elsewhere, here we concentrate on the electronic
structure. Fig.3(b) shows the calculated band structure of LaOFeAs
under pressure (about 30 Gpa). It is clearly seen that the
crossing between the two electron-bands (along M-$\Gamma$ line) is
shifted downward to be less than 1 meV above Fermi level. In the
meanwhile, the size of the electron-type FS is enlarged. All these
indicate the same tendency as discussed above for F-doping, namely
the orbital degeneracy of two electron-type FS is enhanced by
pressure. In addition to the uniform pressure, we also found that
this degeneracy can be enhanced by uniaxial pressure along the $c$
axis. As shown in Fig.3(c), the band crossing is shifted downward
to be about 7meV above Fermi level by 5\% shorting of $c/a$ ratio
with fixed volume.

\begin{figure}[tbp]
\includegraphics[clip,scale=0.35]{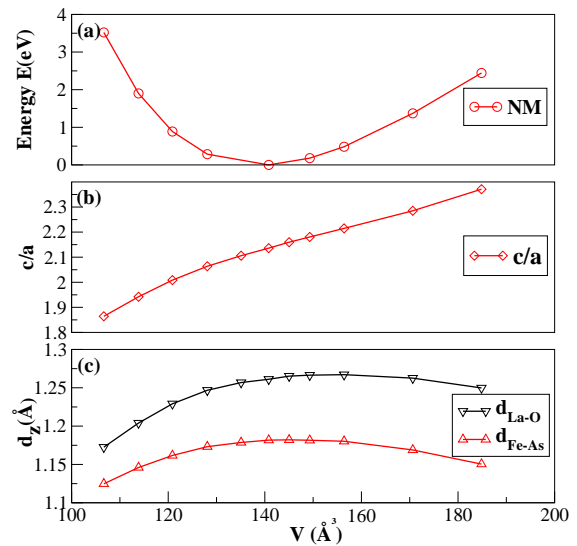}
\caption{The calculated total energy and optimized structure
parameters as function of volume. Here the internal structural
parameter $d_z$ are difined as the inter-layer distance along $c$
axis.}
\end{figure}

The high orbital degeneracy at the FS will lead to strong
inter-orbital fluctuation and may lead to inter-orbital pairing.  In
such case, it will be very easy to understand why the $T_c$ is
enhanced initially by F-doping~\cite{LOFS} and
pressure~\cite{pressure}, and is suppressed beyond the optimal
doping level about 10\%. Based on this inter-band pairing picture,
the possible pairing symmetry has been discussed in our parallel
paper~\cite{Dai}. Here we will show that the experimentally observed
$T_c$ versus F-doping phase diagram can be explained
semi-quantitatively by our scenario. For this purpose, we construct
a minimum two bands effective Hamiltonian to describe the
electron-type FS.

As we have discussed in our parallel paper~\cite{Dai}, the strong
inter-orbital ferromagnetic fluctuation will induce an even parity, orbital
singlet and spin triplet paring state. This tendency will be strongly
enhanced by the large orbital degeneracy at the Fermi surface. Since the
electron pairing happens in momentum space, the main contribution to the
paring comes from the area near the FS. Therefore for the purpose of
illustration, we can use a very simple two band model on the square lattice,
which not only gives the right shape but also the density of states near the
FS for different $F$ doping. The Hamiltonian reads,

\begin{equation}
H=\sum_{k\sigma \alpha }\left( \varepsilon _{k,\alpha }-\mu \right)
C_{k\sigma \alpha }^{\dag }C_{k\sigma \alpha }-J_{k-k^{\prime
}}\sum_{kk^{\prime }m}\widehat{\Delta }_{km}^{\dag }\widehat{\Delta }%
_{k^{\prime }m}
\end{equation}%
where $\alpha =1,2$ are orbital indicies, and

\begin{eqnarray}
\varepsilon _{k,1} &=&t\gamma _{k}+t_{1}\gamma _{k}^{(1)}+t_{2}\gamma
_{k}^{(2)},  \nonumber \\
\varepsilon _{k,2} &=&t\gamma _{k}+t_{2}\gamma _{k}^{(2)}+t_{1}\gamma
_{k}^{(1)},
\end{eqnarray}

with $\gamma _{k}=\cos {k_{x}}+\cos {k_{y},}\gamma _{k}^{(1)}=\cos
(k_{x}+k_{y})$ and $\gamma _{k}^{(2)}=\cos (k_{x}-k_{y})$.

\bigskip By tuning the hopping integrals $t,t_{1}$ and $t_{2}$, we can
obtain very similar FS with the calculated results from
first-principles for different doping. We then solve the above
Hamiltonian by mean field decoupling, as described in detail in our
previous paper~\cite{Dai}. The spacial pairing symmetry is chosen to
be s-wave and k dependence of $J_{k}$ is also ignored for
simplicity. For each doping, we first fit the FS to the GGA results
by tuning the above hopping integrals and then calculate the
super-conducting gap by mean field approach. The results are plotted
in Fig.5. Although our calculation is very rough, it still indicate
two important features of our theory. First, the superconductivity
in the hole doped side and electron doped side is from the same
origin. As pointed out previously~\cite{SDW}, for the pure compound
(i.e. without doping), there is an extra SDW instability due to the
nesting effect between hole and electron FS, which kills the
superconductivity. Both electron and hole doping will remove the FS
nesting effect and superconductivity appears. Second the strongest
superconductivity tendency appears on the electron doping side, with
the doping concentration around 10\%. Both of these predictions seem
quite consistent with the current experimental data. On the other
hand, care must be taken if both F-doping and pressure are applied.
Our prediction is that $T_c$ will be enhanced by pressure for
samples below optimal F-doping (about 10\%), however it is opposite
for overdoped samples.

\begin{figure}[tbp]
\includegraphics[clip,scale=0.45]{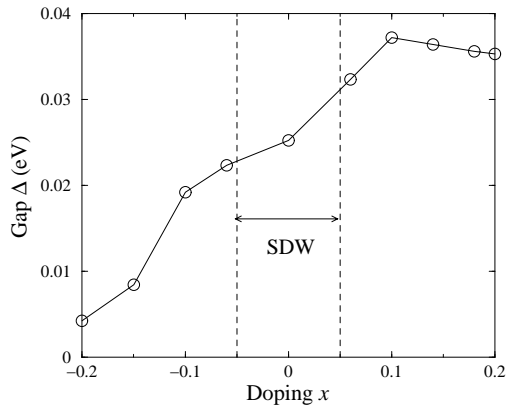}
\caption{The super-conducting gap as the function of doping
concentration for LaO$_{1-x}$F$_{x}$FeAs. The positive $x$ is defined
for electron doping, while negative $x$ for hole-doping. The
inter-band ferromagnetic coupling $J$ is chosen to be 0.45 eV. }
\end{figure}

In summary, by first-principle calculation, we show that the crucial
role played by either F-doping or pressure is to enhance the orbital
degeneracy, which leads to strong inter-band fluctuation. Combining
with experimental observations, we suggest that the inter-orbital
pairing is the key to be considered for the mechanism of
superconductivity in this compound. The same orbital degeneracy is
also present in LaONiAs, and this will be discussed in our separated
paper~\cite{fang-Ni}.

We acknowledge the valuable discussions with Y. P. Wang, F. C. Zhang, and
the supports from NSF of China and that from the 973 program of China
(No.2007CB925000).

\end{document}